\documentclass[prb,twocolumn,showpacs,amsmath,amssymb]{revtex4}

\usepackage{graphicx}
\usepackage{dcolumn}
\usepackage{bm}

\def\mgzno{Zn$_{1-x}$Mg$_x$O}
\def\phm#1{\phantom{#1}}

\begin{document}

\title{First-principles study of polarization in \mgzno}

\author{Andrei Malashevich}
 \email{andreim@physics.rutgers.edu}
\author{David Vanderbilt}
 \email{dhv@physics.rutgers.edu}
\affiliation{
Department of Physics \& Astronomy, Rutgers University,
Piscataway, NJ 08854-8019, USA
}

\date{August 28, 2006}

\begin{abstract}
Wurtzite ZnO can be substituted with up to $\sim$30\% MgO to form a
metastable \mgzno\ alloy while still retaining the wurtzite
structure.  Because this alloy has a larger band gap than pure ZnO,
\mgzno/ZnO quantum wells and superlattices are of interest as
candidates for applications in optoelectronic and
electronic devices.  Here, we report the results of an {\it
ab-initio} study of the spontaneous polarization of \mgzno\ alloys
as a function of their composition.  We perform calculations of the
crystal structure based on density-functional theory in the
local-density approximation, and the polarization is calculated
using the Berry-phase approach.  We decompose the changes in
polarization into purely electronic, lattice-displacement mediated,
and strain mediated components, and quantify the relative importance
of these contributions.   We consider both free-stress and
epitaxial-strain elastic boundary conditions, and show that our
results can be
fairly well reproduced by a simple model in which the piezoelectric
response of pure ZnO is used to estimate the polarization change
of the \mgzno\ alloy induced by epitaxial strain.
\end{abstract}


\maketitle

\section{Introduction}
\label{sec:intro}

Recently, much attention has been paid to wurtzite \mgzno\
alloys as candidates for applications in optoelectronic  devices in
the blue and ultraviolet region. ZnO is a wide-band-gap semiconductor
with a direct gap of $\sim$3.3 eV. The band gap becomes even larger if
Zn atoms are substituted by Mg atoms, which have a similar ionic radius,
allowing the construction of quantum-well and superlattice devices.
\cite{ohtomo}
Similar behavior is well known for the zincblende GaAs/Al$_x$Ga$_{1-x}$As
system and is the basis of much of modern optoelectronics.\cite{adachi}
Recent trends have led in the direction of fabricating similar structures
in wide-gap semiconductor systems such as wurtzite III-V nitrides\cite{bykh}
and in \mgzno.\cite{ohtomo,gruber,zhang} There has also been recent
interest in other kinds of nanostructures based on the ZnO and
\mgzno\ materials systems.\cite{tu,xiang,heo03,heo04}

Pure ZnO prefers the wurtzite crystal structure, while MgO adopts the
cubic rocksalt structure. Substitution of Zn by Mg results in a metastable
wurtzite alloy for certain magnesium concentrations.  Experimental
reports concerning the growth of these alloys on sapphire substrates
indicate that Mg concentrations up to $\sim$30\%,\cite{ohtomo,zhang} or
even $\sim$50\%,\cite{chen} can be achieved.

Many {\it ab-initio} calculations of the properties of the parent
compounds MgO and ZnO have appeared in the
literature.\cite{schleife,corso,limp,gopal}
The properties of ternary \mgzno\ alloys have been less well studied.
There have been calculations of the dependence of the band structure
and band gap on concentration $x$.\cite{lambr}  Regarding the question
of crystal structure and stability,
Kim {\it et al.}\ has shown that the wurtzite \mgzno\ alloy is stable with
respect to the corresponding rocksalt alloy for $x<0.375$.\cite{kim}
Similar results
were obtained by Sanati {\it et al.}\ but for $x<0.33$.\cite{sanati}
However, Sanati {\it et al.}\ also have shown that \mgzno\ is unstable
with respect to phase separation into wurtzite ZnO and rocksalt MgO phases
even for low $x$ values.
This means that \mgzno\ alloys are not thermodynamically stable, consistent
with a rather low observed solid solubility limit for Mg in ZnO.\cite{sarver}
The success in fabricating samples with higher concentrations
indicates that the phase separation is kinetically limited, i.e., the
time scale required for the alloy to phase segregate into the two
lower-energy constituents is long compared to the growth time at the
growth temperature.

To our knowledge, there have not been any previous calculations
of the the polarization properties in the \mgzno\ system.  This is
an important property to study, since if an interface occurs
between a ZnO region and a \mgzno\ region within a superlattice or
quantum-well structure, bound charges are expected to appear at
the interface.  These charges, in turn, will create electric fields
that are likely to affect the electrical and optical properties
of the quantum-well devices.  In the present work, therefore, we
have undertaken a study of the polarization and piezoelectric
properties of \mgzno.

The structure of the paper is as follows. In the next section we
describe the computational methods used in our work. In Sec.~\ref{sec:scell}
we introduce the six supercell structures that were constructed and used
as the structural models for the alloys of interest.
Then, in Sec.~\ref{sec:results}, we report the main results of this
work. Finally, a brief summary is given in Sec.~\ref{sec:summary}.

\section{Computational methods}
\label{sec:methods}

Calculations of structural and polarization properties are
carried out using a plane-wave pseudopotential approach to
density-functional theory (DFT).  We use the ABINIT code package\cite{ABINIT}
with the local-density approximation (LDA) implemented
using the Teter parametrization of the 
exchange-correlation\cite{teter} and with
Troullier-Martins pseudopotentials.\cite{TM}  For the Zn pseudopotential
the $3d$ valence electrons are included in the valence, as their
presence has a significant effect on the accuracy of results.\cite{hill}
A plane-wave basis set with an energy cutoff of 120 Ry is used
to expand the electronic wave functions. A $6\times6\times4$
Brillouin-zone $k$-point sampling is used for pure wurtzite ZnO,
and equivalent $k$-point meshes are constructed for use in all
wurtzite supercell calculations.  The electric polarization is
calculated using the Berry-phase approach.\cite{King-Smith}

\section{Supercell structures}
\label{sec:scell}

In the present work we study the properties of six different models
of the ternary \mgzno\ alloy, to be described shortly.  However,
first consider pure wurtzite ZnO.  It can be viewed as two
identical hexagonal closed-packed ({\it hcp}) lattices; we take
the O sublattice to be shifted in the $+\hat{z}$ direction relative
to the Zn sublattice.
Three parameters determine this structure:  $a$ and
$c$ are the lattice constants of the {\it hcp} lattice, and $u$
describes the shift between the two sublattices.

Replacing some of the Zn atoms by Mg atoms, we get a ternary
\mgzno\ alloy. Of course, the real alloy is highly disordered.  In
order to carry out calculations using periodic boundary conditions,
we construct ordered supercells having the same Mg concentration
$x$ as the alloy of interest.  By comparing properties of different
supercells having the same $x$, we may obtain a rough estimate
of the size of the errors that result from the replacement of the
true disordered alloy by an idealized supercell model.

When constructing supercells, we restricted ourselves to structures
having hexagonal symmetry about the $z$-axis, since real
\mgzno\ alloys have this symmetry on average. This makes the
calculation and interpretation of the results easier.  We
constructed six model alloy structures: one for $x=1/6$ (Model 1),
two for $x=1/4$ (Models 2 and 3), one for $x=1/3$ (Model 4) and two
for $x=1/2$ (Models 5 and 6), as follows.

The simplest alloy one can make (Model 5) is obtained by 
replacing the Zn atoms by Mg atoms in every second Zn layer along
$z$, giving a structure with Mg concentration $x=1/2$ and retaining
the primitive periodicity of pure ZnO (four atoms per cell).
Similarly, if one replaces every fourth layer of Zn by Mg, one
arrives a model with $x=1/4$ (Model 2); this has an eight-atom
supercell with the primitive $1\times1$ in-plane periodicity but
with a doubled periodicity along the $z$-direction.

\begin{figure}
\includegraphics[width=3.4in]{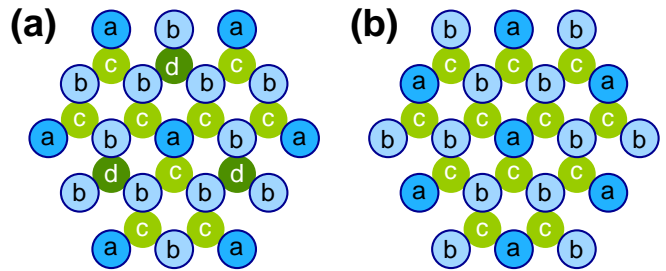}
\caption{\label{fig:mod}
(Color online.) Top view of cation layers of supercell models
for \mgzno\ alloys.
(a) Structures with $2\times2$ periodicity (Models 3 and 6).
(b) Structures with $\sqrt3\times\sqrt3$ periodicity (Models 1 and 4).
Atoms `a' and `b' lie in the top cation layer, while `c' and `d'
are one layer below (see text).}
\end{figure}

In the remaining models, we retain the primitive periodicity along
$z$ but expand the size of the supercell in the $x$-$y$ plane, as
illustrated in Fig.~\ref{fig:mod}.  Models having $2\times2$
in-plane periodicity (Models 3 and 6), and those having
$\sqrt3\times\sqrt3$ periodicity (Models 1 and 4), are specified
with reference to Figs.~\ref{fig:mod}(a) and (b), respectively.
Models 3 and 6 thus have 16 atoms per supercell, while Models 1 and
4 have 12 atoms.  In Model 3 we assign ${\rm b}={\rm c}={\rm Zn}$
and ${\rm a}={\rm d}={\rm Mg}$ in Fig.~\ref{fig:mod}(a), obtaining
a model with $x=1/4$ in each cation layer and $x=1/4$ overall.
Model 6 corresponds to ${\rm b}={\rm d}={\rm Zn}$ and ${\rm a}={\rm
c}={\rm Mg}$; this results in alternating cation layers with
$x=1/4$ and $x=3/4$, for an overall Mg concentration of $x=1/2$.
Turning to the $\sqrt3\times\sqrt3$ structures in
Fig.~\ref{fig:mod}(b), one can see that the hexagonal symmetry
requires that all atoms must be the same ($c$ atoms) in one of the
layers.  We construct Model 1 by assigning ${\rm b}={\rm c}={\rm
Zn}$ and ${\rm a}={\rm Mg}$, yielding alternating layers with $x=0$
and $x=1/3$ for an average $x=1/6$.  Finally, for Model 4 we set ${\rm
a}={\rm c}={\rm Zn}$ and ${\rm b}={\rm Mg}$ so that the layer concentrations
are $x=0$ and $x=2/3$, averaging to $x=1/3$.

Of course, it would be possible to generate more supercell models
of the alloy by expanding the periodicity or reducing the symmetry.
However, the six models described above provide a reasonable
coverage of concentrations in the range $0\le x\le 1/2$ with
some redundancy (for $x=1/4$ and $x=1/2$).  We have thus chosen to
limit ourselves to these six models in the present work.

\section{Results}
\label{sec:results}

\subsection{\label{sec:pure}Pure ZnO and MgO}

To determine the crystal structures and cell parameters of pure ZnO and MgO,
we carried out DFT calculations for both materials in both the
wurtzite and rocksalt
structures. For wurtzite ZnO we obtained lattice parameters $a=3.199$\,\AA,
$c=5.167$\,\AA\ and $u=0.379$. While these results are very close
to previously reported theoretical values,\cite{serrano} they slightly
differ from experimental values\cite{decremps} ($a=3.258$\,\AA,
$c=5.220$\,\AA\ and $u=0.382$).
The cohesive energy (defined as the energy
per formula unit needed to separate the crystal into
atoms) is found to be 8.26\,eV.  Comparing this to
the cohesive energy of rocksalt ZnO (8.03\,eV), one may conclude
that ZnO prefers the wurtzite structure, in agreement with
experiment.  For rocksalt MgO we found $a=4.240$\,\AA\ and a
cohesive energy of 10.00\,eV.
We find that if we start with a plausible wurtzite MgO structure with
$a$, $c$ and $u$ similar to that of ZnO, the crystal can monotonically
lower its energy along a transformation path in which $a$ increases,
$c$ decreases, and $u$ tends toward 1/2 in agreement with
the previous results of Ref.~\onlinecite{limp}. The minimum occurs at
$u=1/2$, which corresponds to the higher-symmetry $h$-MgO structure.
\cite{limp}
For this structure we obtain $a=3.527$\,\AA\ and $c=4.213$\,\AA,
in good agreement\cite{schleife,limp} with previous calculations.
We find its cohesive energy to be 9.81\,eV, consistent with the fact that
MgO prefers the rocksalt structure.  (For more details concerning the previous
theoretical literature on lattice parameters and binding energies,
see Ref.~\onlinecite{schleife}.)

The main goal of the present work is to study the polarization and
piezoelectric properties of \mgzno. For reference, our calculated
spontaneous polarization for pure ZnO is found to be
$-0.0322$\,C/m$^2$, and its piezoelectric coefficients are
$e_{31}=-0.634$\,C/m$^2$ and $e_{33}=1.271$\,C/m$^2$.
Note that the value of the spontaneous polarization differs somewhat
from the previous theory of Dal Corso {\it et al.},\cite{corso} who
reported a polarization of $-0.05$\,C/m$^2$ when using the experimental
$u=0.382$; our value becomes much closer to theirs if we also use the
experimental $u$. 
Since we are primarily interested in {\it differences} of the
polarization with respect to pure ZnO, we do not believe that
these small discrepancies are important.
The values of piezoelectric coefficients are in good agreement with
previous theoretical calculations of Wu {\it et al.}\cite{Xifan}
who found $e_{31}=-0.67$\,C/m$^2$ and $e_{33}=1.28$\,C/m$^2$
(and who also provide comparisons with other theoretical
and experimental results).

\subsection{\label{sec:struct}Crystal structure and energies of alloys}

\begin{table}[t]
\caption{\label{tab:cellp} Theoretical equilibrium lattice parameters
for bulk ZnO and for models of \mgzno.  Subscript `free' indicates
zero-stress elastic boundary conditions, while `epit' indicates
that $a$ is constrained to be identical to that of bulk ZnO
(the values in column V are thus identical by construction).}
\begin{ruledtabular}
\begin{tabular}{llcccc}
 &$x$ &$a_{\rm free}$ (\AA) &${(c/a)}_{\rm free}$ &$a_{\rm epit}$ (\AA)
 &${(c/a)}_{ \rm epit}$\\
\hline
ZnO     & 0.0   & 3.199 & 1.615 & 3.199 & 1.615 \\
Model 1 & 0.17  & 3.216 & 1.605 & 3.199 & 1.624 \\
Model 2 & 0.25  & 3.230 & 1.593 & 3.199 & 1.625 \\
Model 3 & 0.25  & 3.225 & 1.600 & 3.199 & 1.628 \\
Model 4 & 0.33  & 3.238 & 1.589 & 3.199 & 1.630 \\
Model 5 & 0.5   & 3.266 & 1.564 & 3.199 & 1.635 \\
Model 6 & 0.5   & 3.256 & 1.580 & 3.199 & 1.640
\end{tabular}
\end{ruledtabular}
\end{table}

\begin{table}[h]
\caption{\label{tab:cohes} Theoretical cohesive and formation energies
(eV per formula unit) for bulk ZnO and MgO and for each supercell model.}
\begin{ruledtabular}
\begin{tabular}{llrr}
 &$x$ &$E_{\rm coh}$ &$E_{\rm form}$ \\
\hline
ZnO     & 0.0  & 8.258  &    0.0\phm{00} \\
Model 1 & 0.17 & 8.496  & $-$0.053 \\
Model 2 & 0.25 & 8.602  & $-$0.093 \\
Model 3 & 0.25 & 8.612  & $-$0.083 \\
Model 4 & 0.33 & 8.729  & $-$0.123 \\
Model 5 & 0.5  & 8.955  & $-$0.176 \\
Model 6 & 0.5  & 8.958  & $-$0.173 \\
MgO     & 1.0  & 10.004 &    0.0\phm{00}
\end{tabular}
\end{ruledtabular}
\end{table}

For each model described in Sec.~\ref{sec:scell}, we calculated
the {\it hcp} lattice parameters $a$ and $c$ in the equilibrium state.
Since we are interested in properties of \mgzno\ layers that
might be grown on a ZnO substrate, we also calculated
the lattice parameters for epitaxially strained structures (i.e., $a$
fixed to that of pure ZnO).
The results are given in Table \ref{tab:cellp}. In both cases,
the $c/a$ ratio exhibits an almost linear dependence on $x$.
However, this ratio is found to decrease with $x$ for the fully relaxed
structures, while it increases with $x$ when the
epitaxial strain condition is enforced.

In Table \ref{tab:cohes} we give cohesive and formation energies
for each alloy. One can see that in every case the formation energy
is negative.  Thus, according to our LDA calculations, at zero
temperature the \mgzno\ alloy is never stable with respect to
phase-separated wurtzite ZnO and rocksalt MgO.
(Of course, at $T>0$ a small solid solubility of Mg in
wurtzite ZnO is expected.\cite{sarver})

\subsection{\label{sec:polar}Polarization and piezoelectric properties}

The results of the calculations of spontaneous polarization are
given in Table \ref{tab:rel_epit}, both for the fully relaxed
and for the epitaxially strained cases. Note that the values 
of polarization for models having the same $x$ are fairly 
consistent with one another; the choice of supercell does
not significantly affect the overall trend with $x$, which
is reasonably smooth.
A linear fit $P(x)=P({\rm ZnO})+Ax$ yields coefficients of
$A_{\rm free}=(-0.088\pm0.009)$\,C/m$^2$ and
$A_{\rm epit}=( 0.024\pm0.002)$\,C/m$^2$. 
The latter value may be of direct interest for
experimental studies of epitaxial superlattices and quantum wells.

Thus, with
increasing Mg concentration $x$, the absolute value of the
polarization increases for the relaxed structures and decreases
for the epitaxial structures with fixed $a$. This behavior is
very similar to what we saw in Sec.~\ref{sec:struct} for the
$c/a$ ratios, suggesting that the $c/a$ ratio may be a
dominant factor in determining the total polarization. Indeed,
since $2e_{31}+e_{33}\simeq0$, one expects the polarization
to be almost independent of a change in volume (isotropic strain),
so that the change of $c/a$ should be the most important strain effect.

\begin{table}[t]
\caption{\label{tab:rel_epit} Calculated values of total polarizations
of \mgzno\ alloy models (C/m$^2$).  Subscript `free' indicates
zero-stress elastic boundary conditions, while `epit' indicates
that $a$ is constrained to be identical to that of bulk
ZnO. Superscript `est' indicates value estimated by the model
of Eq.~1).}
\begin{ruledtabular}
\begin{tabular}{llrrr}
 &$x$ &$P_{\rm free}$ &$P_{\rm epit}$ &$P_{\rm epit}^{\rm est}$ \\
\hline
ZnO     & 0.0  & $-$0.0322 & $-$0.0322 &           \\
Model 1 & 0.17 & $-$0.0423 & $-$0.0277 & $-$0.0279 \\
Model 2 & 0.25 & $-$0.0501 & $-$0.0247 & $-$0.0247 \\
Model 3 & 0.25 & $-$0.0470 & $-$0.0244 & $-$0.0250 \\
Model 4 & 0.33 & $-$0.0565 & $-$0.0230 & $-$0.0239 \\
Model 5 & 0.5  & $-$0.0789 & $-$0.0199 & $-$0.0222 \\
Model 6 & 0.5  & $-$0.0699 & $-$0.0202 & $-$0.0225
\end{tabular}
\end{ruledtabular}
\end{table}

In order to study more thoroughly the role of strain and other factors
in determining the polarizations of the \mgzno\ structures,
we first define $\Delta P_{\rm tot}$ to be the polarization of the alloy
superlattice structure relative to that of pure ZnO.  We then
decompose $\Delta P_{\rm tot}$ into ``electronic,'' ``ionic,''
and ``piezoelectric'' contributions as follows.  First, we construct
an artificial \mgzno\ superlattice structure in which the structural
paramters
($a$, $c$, and all internal coordinates) are frozen to be those of pure
ZnO, and define $\Delta P_{\rm elec}$ to be the polarization of this
structure relative to that of pure ZnO.  Next, we allow only the
internal coordinates of the \mgzno\ supercell to relax, while continuing
to keep $a$ and $c$ frozen at the pure-ZnO values, and let
$\Delta P_{\rm ion}$ be the polarization change produced by
this internal relaxation.  Finally, we allow the lattice constants to
relax as well, and define $\Delta P_{\rm piezo}$ to be the associated
change in polarization.  Clearly
$\Delta P= \Delta P_{\rm elec}+\Delta P_{\rm ion}+\Delta P_{\rm piezo}$.

The results of such a decomposition are given in Table \ref{tab:contr}
for the stress-free case.
For scale, recall that these are changes relative to
$P$(ZnO)=$-$0.0322\,C/m$^2$.
The purely electronic contributions $\Delta P_{\rm elec}$ are quite
small, showing a relatively poor correlation with $x$.  The contribution
$\Delta P_{\rm ion}$ associated with the ionic relaxations is also
quite small, although it is typically 2-3 times larger than
$\Delta P_{\rm elec}$ and shows a clearer trend (becoming more negative
with increasing $x$).  By far the largest contribution comes from the
piezoelectric effect of the strain relaxation, being typically 5-10 times
larger than the ionic one.
A similar table can be constructed for the case of epitaxial strain;
its first four columns would be identical to
Table \ref{tab:contr} because of the way $\Delta P_{\rm ion}$ and
$\Delta P_{\rm elec}$ are defined, and the values in the remaining
columns can be deduced from the information given in
Tables \ref{tab:rel_epit} and \ref{tab:contr}.  The results indicate
that the piezoelectric contribution also dominates in the
epitaxial-strain case.

This being the case, it
seems likely that many of the polarization-related properties of the
\mgzno\ alloy can be estimated by using a model based on the
piezoelectric effect alone.  For example, one might hope that
$\delta P=P_{\rm epit}-P_{\rm free}$,
the difference between the epitaxially-constrained and free-stress
polarizations at a given $x$, could be estimated
by a linear approximation of the form
\begin{equation}
\delta P=2e_{31}\frac{a_{\rm epit}-a_{\rm free}}{a_{\rm free}}+
e_{33}\frac{c_{\rm epit}-c_{\rm free}}{c_{\rm free}}.
\end{equation}
In fact, we find that this is the case even if we use the piezoelectric
constants of bulk ZnO, already obtained in Sec.~\ref{sec:pure},
in this formula. Using the computed value of $P_{\rm free}$ reported
in the third column of Table \ref{tab:rel_epit}, together with the
constrained $a$ values and epitaxially-relaxed $c$ values given in
the last two columns of Table \ref{tab:cellp}, we report the computed
estimates $P_{\rm epit}^{\rm est}=P_{\rm free}+\delta P$ in the
last column of Table \ref{tab:rel_epit}.
The use of the piezoelectric coefficients of pure ZnO is not obviously
justified except at small $x$, but the results show excellent agreement
with the computed $P_{\rm epit}$ values in the fourth column even
up to $x=0.5$, where the error is only about 10\%.  This approximation
thus seems to work quite well.

\begin{table}[t]
\caption{\label{tab:contr} Theoretical values of electronic, ionic,
piezoelectric and total contributions to polarization (C/m$^2$) for each
model, relative to bulk ZnO.}
\begin{ruledtabular}
\begin{tabular}{llrrrr}
 &$x$ &$\Delta P_{\rm elec}$ &$\Delta P_{\rm ion}$ &$\Delta P_{\rm piezo}$ &$\Delta P_{\rm tot}$\\
\hline
ZnO     & 0.0  & 0.0\phm{000} & 0.0\phm{000} & 0.0\phm{000} & 0.0\phm{000} \\
Model 1 & 0.17 &    0.0001 & $-$0.0022 & $-$0.0081 & $-$0.0101 \\
Model 2 & 0.25 &    0.0018 & $-$0.0023 & $-$0.0175 & $-$0.0180 \\
Model 3 & 0.25 &    0.0000 & $-$0.0027 & $-$0.0122 & $-$0.0148 \\
Model 4 & 0.33 &    0.0009 & $-$0.0038 & $-$0.0214 & $-$0.0243 \\
Model 5 & 0.5  &    0.0023 & $-$0.0063 & $-$0.0427 & $-$0.0467 \\
Model 6 & 0.5  & $-$0.0019 & $-$0.0062 & $-$0.0296 & $-$0.0377
\end{tabular}
\end{ruledtabular}
\end{table}
\section{Summary}
\label{sec:summary}

We have investigated the polarization-related properties of
wurtzite \mgzno\ alloys using calculations based on
density-functional theory in the local-density approximation and
the Berry-phase approach to calculating electric polarization.  In
particular, we have studied the dependence of the spontaneous
polarization on Mg concentration using six alloy supercell models
with hexagonal symmetry,
spanning the range of Mg concentration from $x=1/6$ to $1/2$.
We performed these calculations both for free-stress and
epitaxial-strain elastic boundary conditions.

Our results indicate
a roughly linear dependence of spontaneous polarization on Mg
concentration, although the sign of the linear coefficient is
opposite in the free-stress and epitaxial-strain cases.  In order
to understand this behavior in more detail, we decomposed the
change in polarization into electronic,
lattice-displacement-mediated, and strain-mediated components, and
found that the latter component is dominant.  This means that the
change in polarization is mostly governed by piezoelectric effects
connected with the $x$-dependent changes of the $a$ and $c$ lattice
constants.  We further confirmed this picture by showing that the
polarization changes could be well approximated by a model
in which the only first-principles inputs to the model are the
piezoelectric coefficients of pure ZnO and the $x$-dependence of
the equilibrium lattice constants of the \mgzno\ alloy.
These results suggest that charging effects associated with
polarization discontinuities in ZnO/\mgzno\ superlattices and quantum
wells should be subject to prediction and interpretation in a fairly
straightforward manner.


\acknowledgments

This work was supported by NSF Grant DMR-0549198.

\bibliography{paper}

\end{document}